\documentclass[twocolumn,showpacs,preprintnumbers]{revtex4}%
\usepackage{amssymb}
\usepackage{amsfonts}
\usepackage{amsmath}
\usepackage{graphicx}
\usepackage{dcolumn}
\usepackage{bm}%
\begin{document}
\preprint{ }
\title{Unexpected Nonlinear Dynamics in NbN Superconducting Microwave Resonators}
\author{B. Abdo}
\email{baleegh@tx.technion.ac.il}
\author{E. Segev}
\author{O. Shtempluck}
\author{E. Buks}
\affiliation{Microelectronics Research Center, Department of Electrical Engineering,
Technion, Haifa 32000, Israel}
\date{\today}

\begin{abstract}
In this work we characterize the unusual nonlinear dynamics of the resonance
response, exhibited by our NbN superconducting microwave resonators, using
different operating conditions. The nonlinear dynamics, occurring at
relatively low input powers (2-4 orders of magnitude lower than Nb), and which
include among others, bifurcations in the resonance curve, hysteresis loops
and resonance frequency shift, are measured herein using varying temperature,
applied magnetic field, white noise and rapid frequency sweeps. Based on these
measurement results, we consider a hypothesis according to which Josephson
junctions forming weak links at the boundaries of the NbN grains are
responsible for the observed behavior, and we show that most of the
experimental results are qualitatively consistent with such hypothesis.

\end{abstract}
\maketitle



%It is always \today, today,
%but any date may be explicitly specified

\section{INTRODUCTION}

Understanding the underlying mechanisms that cause and manifest nonlinear
effects in superconductors has a significant importance both theoretically and
practically. On the one hand, knowing how to minimize and control nonlinear
effects, being the major cause for the degradation of performance of RF
devices employing superconductors \cite{RF power dependence}, would have a
great technological impact. It will enable the design and fabrication of high
performance superconducting RF devices with extended linear behavior, and
higher power densities than afforded nowadays \cite{Intrinsic,nonlinear
dynamics}; On the other hand, identifying the origins of the nonlinearities in
superconductors, would lead to a better understanding of the superconducting
phenomena and its behavior in the microwave regime.

In spite of the intensive study of nonlinearities in superconductors in the
past decades, we still lack a coherent picture regarding these effects and
their origins \cite{understanding}. This is partly because the nonlinear
mechanisms in superconductors are various and many times act concurrently on
the superconducting sample, hence making the identification process of the
dominant factor mainly indirect and based on eliminations \cite{Jerusalem}.

Nonlinear mechanisms in superconductors, which are usually divided into
intrinsic and extrinsic origins, include among others, Meissner effect
\cite{Yip}, pair-breaking, global and local heating effects \cite{Thermally
induced nonlinear behaviour,Thermally-induced nonlinearities surface impedance
sc YBCO}, rf and dc vortex penetration and motion \cite{vortices}, defect
points, damaged edges \cite{edge}, substrate material \cite{anomalies in
nonlinear mw surface vs sub effects}, and weak links \cite{Halbritter}.
Whereas weak links is a collective term representing various material defects
located inside the superconductor such as weak superconducting points
switching to normal state under low current density, Josephson junctions
forming inside the superconductor structure, grain-boundaries, voids,
insulating oxides, insulating planes. These defects and impurities generally
affect the conduction properties of the superconductor and as a result cause
extrinsic nonlinear effects.

Nonlinear effects in superconductors in general and in NbN in particular have
been reported and analyzed by several research groups. Duffing like
nonlinearity for example was observed in different superconducting resonators
employing different geometries and materials. It was observed in a
HTS\ parallel plate resonator \cite{Thermally induced nonlinear behaviour}, in
a Nb microstrip resonator \cite{microwave nonlinear effects in He-cooled sc
microstrip resonators}, in a Nb \ and NbN stripline resonators \cite{nonlinear
dynamics}, in a YBCO coplanar-waveguide resonator \cite{RF power dependence CP
YBCO}, in a YBCO thin film dielectric cavity \cite{Thermally-induced
nonlinearities surface impedance sc YBCO}, and also in a suspended HTS thin
film resonator \cite{suspended HTS mw resonator}. Other nonlinearities
including notches, anomalies developing at the resonance lineshape and
frequency hysteresis can be found in \cite{HTS patch antenna,power dependent
effects observed for sc stripline resonator}.

In this paper we report the observation of unexpected nonlinear dynamics
measured in NbN superconducting microwave resonators. Among the observed
effects, asymmetric resonances, multiple bifurcations in the resonance
lineshape, hysteretic behavior in the vicinity of the bifurcations, hysteresis
loops changing direction, resonance frequency shift as the input power is
increased and nonlinear coupling. Some of these nonlinear effects were
introduced in a previous publication \cite{nonlinear features BB}, thus this
paper will concentrate on presenting a different set of measurements applied
to these nonlinear resonators, which provides a better understanding of the
underlying physical mechanism causing these effects. For this purpose, we have
measured the nonlinear superconducting resonators under different operating
conditions, such as added white noise (to the main signal), fast frequency
sweep using frequency modulation (FM), applied magnetic field, and varying
temperature. Under each one of these operating conditions we observe
interesting nonlinear dynamics, which qualitatively agree with our hypothesis
of microscopic Josephson junctions forming at the boundaries of the NbN
columnar structure.

The remainder of this paper is organized as follows, the fabrication process
of the NbN superconducting resonators is described briefly in Sec. II. A short
summary of previous results is brought in Sec. III. The nonlinear dynamics of
these microwave resonators measured under various operating conditions are
reviewed in Sec. IV. Whereas possible underlying physical mechanisms
responsible for the measured effects are discussed in Sec. V, which also
concludes this paper.

\section{FABRICATION PROCESS}

\begin{figure}
[ptb]
\begin{center}
\includegraphics[
height=1.5108in,
width=3.1661in
]%
{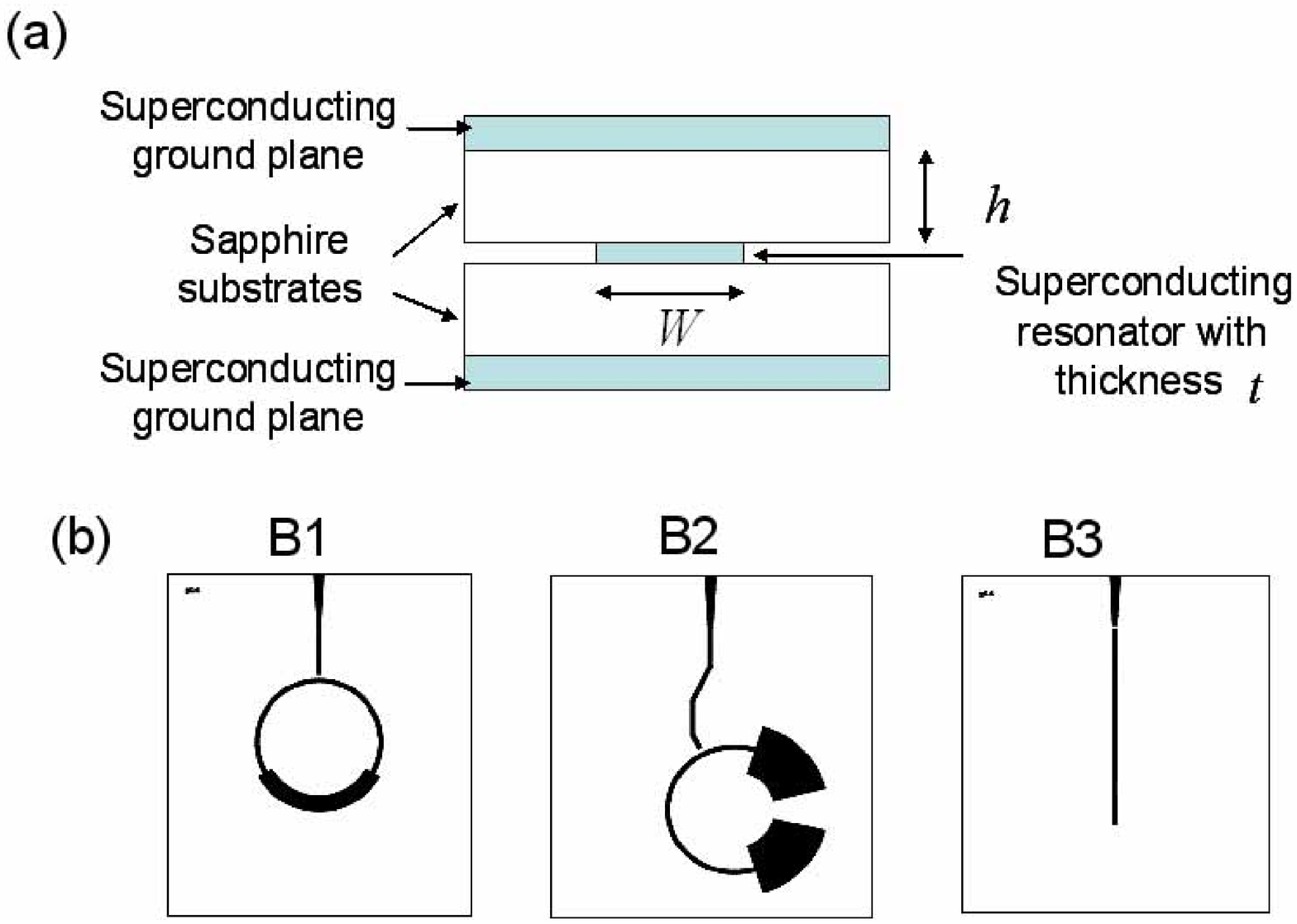}%
\caption{(a) Schematic cross section of the stripline resonator used,
consisting of five layers: two superconducting ground planes, two sapphire
substrates and a NbN film in the middle deposited on one of the sapphires. (b)
Top view of the three resonator layouts (B1, B2, B3) which were deposited as
the middle layer. }%
\label{layout}%
\end{center}
\end{figure}

The measurement results presented in this paper belong to three nonlinear NbN
superconducting microwave resonators. The resonators were fabricated using
stripline geometry, consisting of two superconducting ground planes, two
sapphire substrates, and a deposited strip in the middle (the deposition was
done on one of the sapphire substrates). Fig. \ref{layout} shows a schematic
diagram illustrating stripline geometry and a top view of the three resonators
layouts. For convenience we will be referring to the three resonators in the
text by the shortened names B1, B2 and B3 as defined in Fig. \ref{layout}. The
sapphire substrates dimensions used were $34%
%TCIMACRO{\unit{mm}}%
%BeginExpansion
\mathrm{mm}%
%EndExpansion
$ X $30%
%TCIMACRO{\unit{mm}}%
%BeginExpansion
\mathrm{mm}%
%EndExpansion
$ X $1%
%TCIMACRO{\unit{mm}}%
%BeginExpansion
\mathrm{mm}%
%EndExpansion
,$ whereas the coupling gap between the resonators and their feedline was set
to $0.4%
%TCIMACRO{\unit{mm}}%
%BeginExpansion
\mathrm{mm}%
%EndExpansion
$ in B1 and B3 cases and $0.5%
%TCIMACRO{\unit{mm}}%
%BeginExpansion
\mathrm{mm}%
%EndExpansion
$ in B2 resonator. The resonators were dc-magnetron sputtered in a mixed
Ar/N$_{2}$ atmosphere, near room temperature. The patterning was done using
standard UV photolithography process, whereas the NbN etching was performed by
Ar ion-milling. The sputtering parameters, design consideration as well as
other fabrication details can be found elsewhere \cite{nonlinear features BB}.
The critical temperature $T_{c}$ of B1, B2 and B3 resonators were relatively
low and equal to $10.7%
%TCIMACRO{\unit{K}}%
%BeginExpansion
\mathrm{K}%
%EndExpansion
,$ $6.8%
%TCIMACRO{\unit{K}}%
%BeginExpansion
\mathrm{K}%
%EndExpansion
$ and $8.9%
%TCIMACRO{\unit{K}}%
%BeginExpansion
\mathrm{K}%
%EndExpansion
$ respectively. The thickness of the NbN resonators were $2200$ $%
%TCIMACRO{\unit{\U{212b}}}%
%BeginExpansion
\mathrm{\text{\AA}}%
%EndExpansion
$ in B1, $3000$ $%
%TCIMACRO{\unit{\U{212b}}}%
%BeginExpansion
\mathrm{\text{\AA}}%
%EndExpansion
$ in B2, and $2000$ $%
%TCIMACRO{\unit{\U{212b}}}%
%BeginExpansion
\mathrm{\text{\AA}}%
%EndExpansion
$ in B3 resonator.%

%TCIMACRO{\FRAME{ftbpFU}{3.1661in}{1.5108in}{0pt}{\Qcb{(a) Schematic cross
%section of the stripline resonator used, consisting of five layers: two
%superconducting ground planes, two sapphire substrates and a NbN film in the
%middle deposited on one of the sapphires. (b) Top view of the three resonator
%layouts (B1, B2, B3) which were deposited as the middle layer. }}%
%{\Qlb{layout}}{layout.eps}{\special{ language "Scientific Word";
%type "GRAPHIC";  display "USEDEF";  valid_file "F";  width 3.1661in;
%height 1.5108in;  depth 0pt;  original-width 8.3212in;
%original-height 5.7294in;  cropleft "0";  croptop "1";  cropright "1";
%cropbottom "0";  filename 'layout.eps';file-properties "XNPEU";}}}%
%BeginExpansion
%EndExpansion

\section{SUMMARY OF PREVIOUS RESULTS}

\begin{figure}
[ptb]
\begin{center}
\includegraphics[
height=2.4146in,
width=3.5924in
]%
{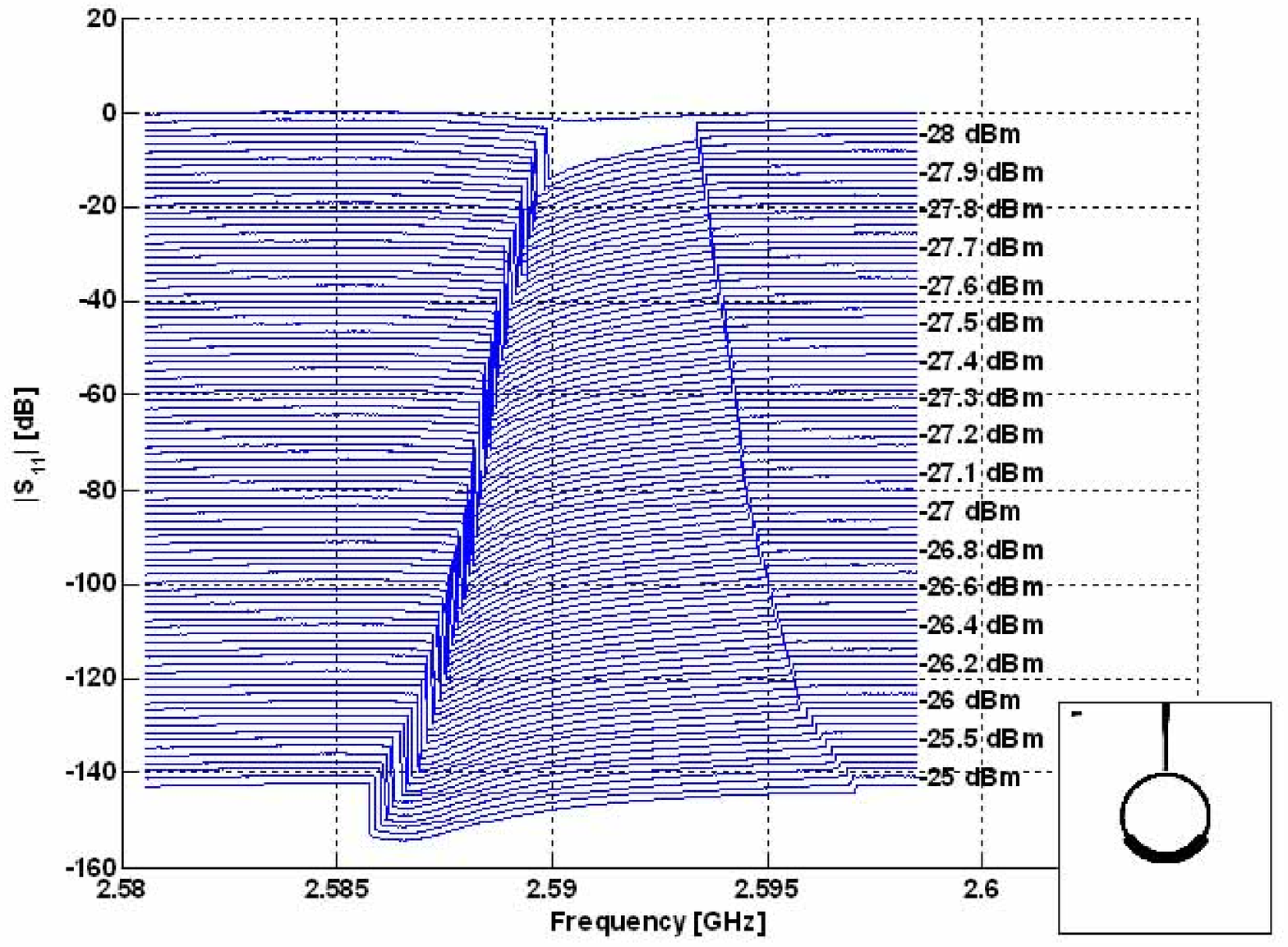}%
\caption{$S_{11}$ amplitude measurement of B1 resonator at its first mode
$\sim2.59\mathrm{GHz}.$ The measured resonance lineshapes are
asymmetrical and contain two abrupt bifurcations at the sides of the
resonance. Moreover the bifurcation frequencies shift outwards as the input
power is increased. The measured resonance lineshapes were shifted vertically
by a constant offset for clarity. The layout of B1 resonator is depicted in
the inset.}%
\label{b1s11f1}%
\end{center}
\end{figure}

As we have shown comprehensively in a previous publication \cite{nonlinear
features BB}, two of the main nonlinear features observed in our NbN nonlinear
superconducting resonators are bifurcations that appear in the resonance
curves at relatively low input powers and hysteresis loops forming in the
vicinity of the bifurcations. In Fig. \ref{b1s11f1} we present a $S_{11\text{
}}$parameter measurement of B1 first mode resonance featuring bifurcations at
both sides of the resonance curve.%

%TCIMACRO{\FRAME{ftbpFU}{3.5924in}{2.4146in}{0pt}{\Qcb{$S_{11}$ amplitude
%measurement of B1 resonator at its first mode $\sim2.59\unit{GHz}.$ The
%measured resonance lineshapes are asymmetrical and contain two abrupt
%bifurcations at the sides of the resonance. Moreover the bifurcation
%frequencies shift outwards as the input power is increased. The measured
%resonance lineshapes were shifted vertically by a constant offset for clarity.
%The layout of B1 resonator is depicted in the inset.}}{\Qlb{b1s11f1}%
%}{b1s11f1.eps}{\special{ language "Scientific Word";  type "GRAPHIC";
%display "USEDEF";  valid_file "F";  width 3.5924in;  height 2.4146in;
%depth 0pt;  original-width 10.0145in;  original-height 7.5057in;
%cropleft "0";  croptop "1";  cropright "1";  cropbottom "0";
%filename 'b1s11f1.eps';file-properties "XNPEU";}}}%
%BeginExpansion
%EndExpansion

\begin{figure}
[ptb]
\begin{center}
\includegraphics[
height=1.5904in,
width=2.2494in
]%
{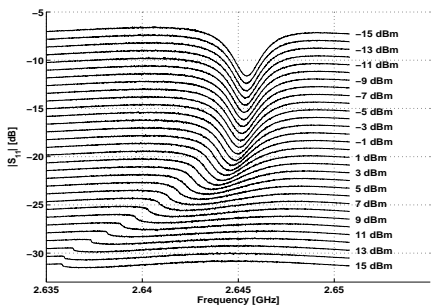}%
\caption{Duffing nonlinearity exhibited by a Nb stripline resonator employing
B2 layout at its first mode. The different $S_{11}$ amplitude plots correspond
to different input powers, ranging from $-15$ dBm to $15$ dBm in steps of $2$
dBm. As the input power is increased the resonance becomes asymmetrical and
infinite slope develops at the left side of the resonance curve. The plots
were offset in the vertical direction for clarity.}%
\label{nb_duffing}%
\end{center}
\end{figure}

\begin{figure}
[ptb]
\begin{center}
\includegraphics[
height=2.4189in,
width=3.5008in
]%
{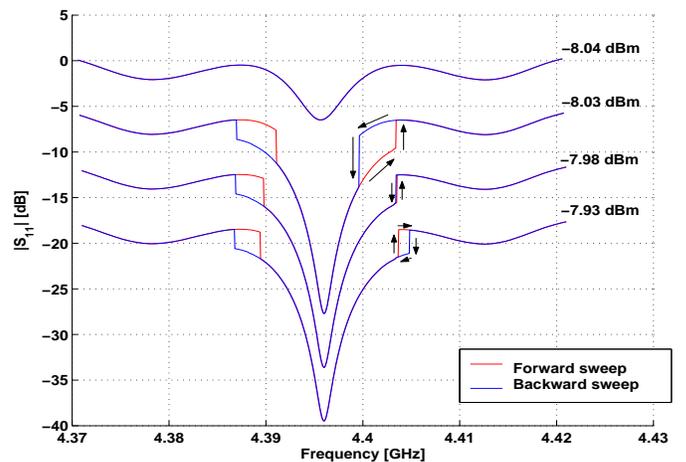}%
\caption{Frequency sweep measurement of B2 resonator at $\sim
4.395\mathrm{GHz}$ performed in both frequency directions. The plots
exhibit hysteresis loops forming at the vicinity of the bifurcations, as well
as hysteresis loops changing direction as the input power is increased. The
red line represents a forward sweep, whereas the blue line represents a
backward sweep. The resonance lineshapes were shifted vertically by a constant
offset for clarity.}%
\label{hysteresis}%
\end{center}
\end{figure}

Furthermore, the frequencies at which the bifurcations occur shift outwards
from the center frequency as the input power is increased. The measurement was
performed, at liquid helium temperature $4.2%
%TCIMACRO{\unit{K}}%
%BeginExpansion
\mathrm{K}%
%EndExpansion
,$ using vector network analyzer. The measured curves in the figure were
shifted vertically by a constant offset for clarity. In contrast to the
nonlinear dynamics presented in Fig. \ref{b1s11f1}, we show in Fig.
\ref{nb_duffing}, a resonance response measured at $4.2%
%TCIMACRO{\unit{K}}%
%BeginExpansion
\mathrm{K}%
%EndExpansion
,$ exhibiting Duffing nonlinearity of the kind generally reported in the
literature \cite{nonlinear dynamics,RF power dependence CP YBCO,suspended HTS
mw resonator}. This nonlinearity which can be explained in terms of resistance
change $\Delta R$ and kinetic inductance change $\Delta L_{K}$
\cite{Dahm,Nonlinear TL} was measured at the first resonance frequency of a
$2200$ $%
%TCIMACRO{\unit{\U{212b}}}%
%BeginExpansion
\mathrm{\text{\AA}}%
%EndExpansion
$ thickness Nb resonator employing B2 layout geometry ($T_{c}=8.9%
%TCIMACRO{\unit{K}}%
%BeginExpansion
\mathrm{K}%
%EndExpansion
$). The differences between the two nonlinear dynamics are obvious, but
nevertheless note the difference in the order of magnitude of the input powers
at which these two nonlinearities occur, $\sim-28$ dBm in the NbN case vs.
$\sim10$ dBm in the Nb resonator. In Fig. \ref{hysteresis} we present a
$\ S_{11\text{ }}$measurement of B2 resonator at its second mode, measured
while sweeping the frequency in both directions. The red line represents a
forward frequency sweep whereas the blue line represents a backward frequency
sweep. The graphs corresponding for different input powers were also offset in
the vertical direction for clarity. At $-8.04$ dBm the resonance is linear and
shows no hysteresis behavior. As the input power is increased by a power step
of $0.01$ dBm, to $-8.03$ dBm the resonance response lineshape is changed
dramatically, featuring two bifurcations at both sides of the resonance curve
and hysteresis loops forming in the vicinity of the bifurcations (as can be
clearly seen in the figure). Moreover this measurement exhibits another
interesting nonlinear feature. The right side hysteresis loop changes
direction as the input power is increased. Between $-8.03$ dBm and $-7.98$ dBm
the hysteresis loop circulates counter clockwise. At $-7.98$ dBm the
hysteresis loop vanishes. Whereas as the input power is increased further, the
loop circulates clockwise. Furthermore at $-6.35$ dBm (not shown here) the
hysteresis loop vanishes again and changes its direction one more time, (it
starts circulating counter clockwise as the power is increased).

\begin{figure}
[ptb]
\begin{center}
\includegraphics[
height=2.5382in,
width=3.3814in
]%
{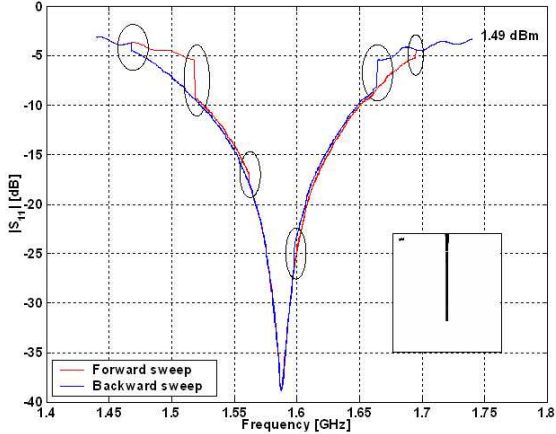}%
\caption{$S_{11}$ parameter amplitude measurement of the first resonance of B3
resonator, (shown in the inset), measured at input power of $1.49$ dBm. The
measurement was done using a network analyzer employing $4000$ measurement
points in each direction. The red line represents a forward frequency scan
whereas the blue line represents a backward scan. The plot shows clearly three
bifurcations within the resonance lineshape in each direction, as indicated by
small circles. }%
\label{b3fb}%
\end{center}
\end{figure}

In Fig. \ref{b3fb} we show yet another nonlinear feature exhibited by these
nonlinear resonators, namely multiple bifurcations in the resonance curve. The
figure plots the resonance response of B3 resonator at its first resonance
frequency, corresponding to $1.49$ dBm input power. The measurement which was
obtained by sweeping the frequency axis in the forward and backward
directions, features three bifurcations and hysteresis loops within the
resonance lineshape in each direction. 

Other experimental data featuring these effects and similar nonlinearities
observed in B1, B2 and B3 resonators can be found in Ref. \cite{nonlinear
features BB}. Among the nonlinearities not brought here or in \cite{nonlinear
features BB}, one can name, nonlinear coupling \cite{nonlinear coupling}, and
intermodulation gain \cite{IMD amplifier}.%

%TCIMACRO{\FRAME{ftbpFU}{2.2494in}{1.5904in}{0pt}{\Qcb{Duffing nonlinearity
%exhibited by a Nb stripline resonator employing B2 layout at its first mode.
%The different $S_{11}$ amplitude plots correspond to different input powers,
%ranging from $-15$ dBm to $15$ dBm in steps of $2$ dBm. As the input power is
%increased the resonance becomes asymmetrical and infinite slope develops at
%the left side of the resonance curve. The plots were offset in the vertical
%direction for clarity.}}{\Qlb{nb_duffing}}{nb_duffing.eps}%
%{\special{ language "Scientific Word";  type "GRAPHIC";  display "USEDEF";
%valid_file "F";  width 2.2494in;  height 1.5904in;  depth 0pt;
%original-width 6.9444in;  original-height 5.4942in;  cropleft "0";
%filename 'nb_duffing.eps';file-properties "XNPEU";}}}%
%BeginExpansion
%EndExpansion

%

%TCIMACRO{\FRAME{ftbpFU}{3.5008in}{2.4189in}{0pt}{\Qcb{Frequency sweep
%measurement of B2 resonator at $\sim4.395\unit{GHz}$ performed in both
%frequency directions. The plots exhibit hysteresis loops forming at the
%vicinity of the bifurcations, as well as hysteresis loops changing direction
%as the input power is increased. The red line represents a forward sweep,
%whereas the blue line represents a backward sweep. The resonance lineshapes
%were shifted vertically by a constant offset for clarity.}}{\Qlb{hysteresis}%
%}{hysteresis.eps}{\special{ language "Scientific Word";  type "GRAPHIC";
%display "USEDEF";  valid_file "F";  width 3.5008in;  height 2.4189in;
%depth 0pt;  original-width 7.1131in;  original-height 5.4942in;
%cropleft "0";  croptop "1";  cropright "1";  cropbottom "0";
%filename 'hysteresis.eps';file-properties "XNPEU";}}}%
%BeginExpansion
%EndExpansion
%

%TCIMACRO{\FRAME{ftbpFU}{3.3814in}{2.5382in}{0pt}{\Qcb{$S_{11}$ parameter
%amplitude measurement of the first resonance of B3 resonator, (shown in the
%inset), measured at input power of $1.49$ dBm. The measurement was done using
%a network analyzer employing $4000$ measurement points in each direction. The
%red line represents a forward frequency scan whereas the blue line represents
%a backward scan. The plot shows clearly three bifurcations within the
%resonance lineshape in each direction, as indicated by small circles. }%
%}{\Qlb{b3fb}}{b3fb.eps}{\special{ language "Scientific Word";
%type "GRAPHIC";  maintain-aspect-ratio TRUE;  display "USEDEF";
%valid_file "F";  width 3.3814in;  height 2.5382in;  depth 0pt;
%original-width 10.0024in;  original-height 7.4936in;  cropleft "0";
%filename 'B3fb.eps';file-properties "XNPEU";}}}%
%BeginExpansion
%EndExpansion

\section{MEASUREMENT RESULTS}

In the following subsections we present measurement results of B1 and B2
resonators, under different operating conditions. In subsection A, the impact
of white noise power on the bifurcation feature is briefly examined. The
dependence of the nonlinear dynamics on temperature varying and applied
magnetic field are investigated in subsections B and C respectively. In
subsection D, heating effects are considered through frequency sweep time
analysis and comparisons to heating time scales previously reported in the
literature. Evidence of the columnar structure of our NbN films is presented
in subsection E. Possible interpretations of these observations are discussed
in Sec. V.

\subsection{WHITE NOISE EFFECT}

\begin{figure}
[ptb]
\begin{center}
\includegraphics[
height=1.0499in,
width=2.5114in
]%
{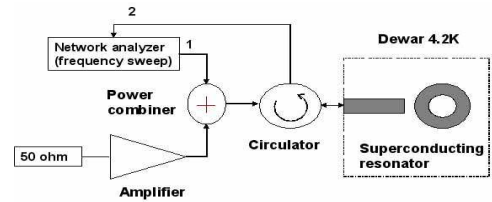}%
\caption{Schematic diagram of the experimental setup used to measure the
nonlinear resonance $2.585$ $\mathrm{GHz}$ of B1 using frequency sweep
mode while applying $-58$ dBm white noise. }%
\label{noisesetup}%
\end{center}
\end{figure}

\begin{figure}
[ptb]
\begin{center}
\includegraphics[
height=2.2364in,
width=3.563in
]%
{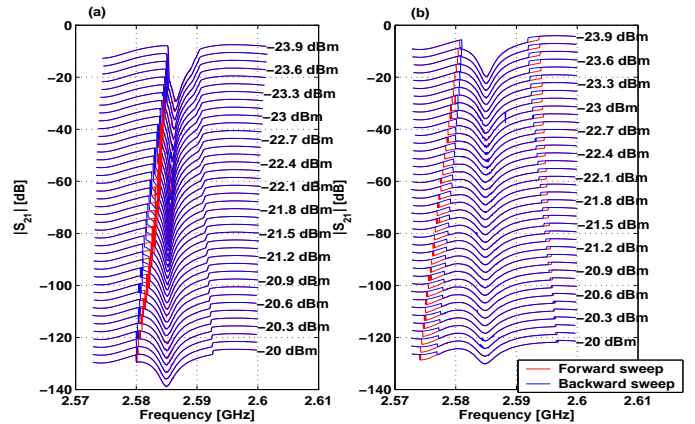}%
\caption{Frequency sweep measurement of B1 resonator at $2.585$
$\mathrm{GHz}$ performed in both directions (a) while applying white
noise of $-58$ dBm/Hz (b) without applying external noise ($\sim-80$
dBm/$\mathrm{Hz}$). The red line represents a forward sweep, whereas the
blue line represents a backward sweep. The measured resonance curves were
shifted vertically by a constant offset for clarity.}%
\label{s21compnoise}%
\end{center}
\end{figure}

Bifurcations in the resonance response of nonlinear oscillators is usually
described in terms of metastable states and dynamic transition between basins
of attraction of the oscillator \cite{Cleland}. Thus in order to examine
qualitatively the metastability of the resonance bifurcations of these
nonlinear resonators, we have applied a constant white noise power to the
resonator, several orders of magnitude lower than the main signal power, using
the setup shown in Fig. \ref{noisesetup}. The white noise level applied was
$-58$ dBm/Hz. It was generated by amplifying the thermal noise of a room
temperature $50$ $%
%TCIMACRO{\unit{\U{3a9}}}%
%BeginExpansion
\mathrm{\Omega }%
%EndExpansion
$ load using an amplifying stage. The generated noise was added to the
transmitted power of the network analyzer via a power combiner. The power
reflections were redirected by a circulator on their way back, and were
measured at the second port of the network analyzer, thus measuring $S_{21}$
parameter. The effect of the $-58$ dBm/Hz white noise power on B1 first mode
bifurcations, is shown in Fig. \ref{s21compnoise} (a), whereas for comparison
in Fig. \ref{s21compnoise} (b) we show nearly noiseless ($\sim-80$ dBm/Hz)
resonance curve measurements performed using the simple setup depicted in Fig.
\ref{noisesetup} without the amplifier and the combiner stage. The two
measurements were carried out within the same input power range ($-23.9$ dBm
through $-20$ dBm).

By comparing between the two measurement results, one can observe the
following, the two fold bifurcations in Fig. \ref{s21compnoise} (b) form a
hysteresis loop at each side of the resonance curve. In Fig.
\ref{s21compnoise} (a) in contrast, as a result of the added noise, the
bifurcations at the left side, represented by the thick colored lines on the
graphs, become frequent and bidirectional, whereas the hysteresis loops, at
the right side, vanish.

The transition rate $\Gamma\left(  f\right)  $ between the oscillator basins
of attraction (as a function the oscillator frequency $f$), can be generally
estimated by the expression \cite{Cleland} $\Gamma\left(  f\right)
=\Gamma_{0}\exp\left(  -E_{A}\left(  f\right)  /k_{B}T_{eff}\right)  ,$ where
$E_{A}\left(  f\right)  $ is the quasi-activation energy of the oscillator,
$T_{eff}$ is proportional to the noise power, $k_{B}$ is Boltzmann's constant,
$f$ is the oscillator frequency, whereas $\Gamma_{0}$ is related to Kramers
low-dissipation form \cite{Kramer} and it is given approximately by $f_{0}/Q$,
where $f_{0}$ is the natural resonance frequency, and $Q$ is the quality
factor of the oscillator. Based on the results presented in Fig.
\ref{s21compnoise}, one can evaluate the following parameters $f_{0}%
\simeq2.585%
%TCIMACRO{\unit{GHz}}%
%BeginExpansion
\mathrm{GHz}%
%EndExpansion
,$ $Q\sim1620,$ $\Gamma_{0}\sim1.6%
%TCIMACRO{\unit{MHz}}%
%BeginExpansion
\mathrm{MHz}%
%EndExpansion
,$ $T_{eff}\sim10^{14}%
%TCIMACRO{\unit{K}}%
%BeginExpansion
\mathrm{K}%
%EndExpansion
.$ Whereas $E_{A}$ the quasi-activation energy of the oscillator may be
roughly estimated for the left and right sides of the resonance response
separately. For the right side of the resonance curve, the quenching of the
hysteresis as a result of the white noise, suggests that the applied noise
power $-58$ dBm/$%
%TCIMACRO{\unit{Hz}}%
%BeginExpansion
\mathrm{Hz}%
%EndExpansion
$ is within the high noise limit, where the noise power is higher than the
energy barrier associated with this bifurcation. Whereas for the left side,
the hysteresis loop and the noise-induced transitions, indicate that the noise
power $-58$ dBm/$%
%TCIMACRO{\unit{Hz}}%
%BeginExpansion
\mathrm{Hz}%
%EndExpansion
$ is still in the moderate noise limit, where the\ noise power\ is less than
the quasi-activation energy associated with this bifurcation.%

%TCIMACRO{\FRAME{ftbpFU}{2.5114in}{1.0499in}{0pt}{\Qcb{Schematic diagram of the
%experimental setup used to measure the nonlinear resonance $2.585$
%$\unit{GHz}$ of B1 using frequency sweep mode while applying $-58$ dBm white
%noise. }}{\Qlb{noisesetup}}{noisesetup.eps}%
%{\special{ language "Scientific Word";  type "GRAPHIC";  display "USEDEF";
%valid_file "F";  width 2.5114in;  height 1.0499in;  depth 0pt;
%original-width 9.7222in;  original-height 4.2627in;  cropleft "0";
%filename 'noisesetup.eps';file-properties "XNPEU";}}}%
%BeginExpansion
%EndExpansion
%

%TCIMACRO{\FRAME{ftbpFU}{3.563in}{2.2364in}{0pt}{\Qcb{Frequency sweep
%measurement of B1 resonator at $2.585$ $\unit{GHz}$ performed in both
%directions (a) while applying white noise of $-58$ dBm/Hz (b) without applying
%external noise ($\sim-80$ dBm/$\unit{Hz}$). The red line represents a forward
%sweep, whereas the blue line represents a backward sweep. The measured
%resonance curves were shifted vertically by a constant offset for
%clarity.}}{\Qlb{s21compnoise}}{s21compnoise.eps}%
%{\special{ language "Scientific Word";  type "GRAPHIC";  display "USEDEF";
%valid_file "F";  width 3.563in;  height 2.2364in;  depth 0pt;
%original-width 10.6666in;  original-height 6.8822in;  cropleft "0";
%filename 's21compnoise.eps';file-properties "XNPEU";}}}%
%BeginExpansion
%EndExpansion

\subsection{TEMPERATURE DEPENDENCE}

\begin{figure}
[ptb]
\begin{center}
\includegraphics[
height=2.5512in,
width=3.8623in
]%
{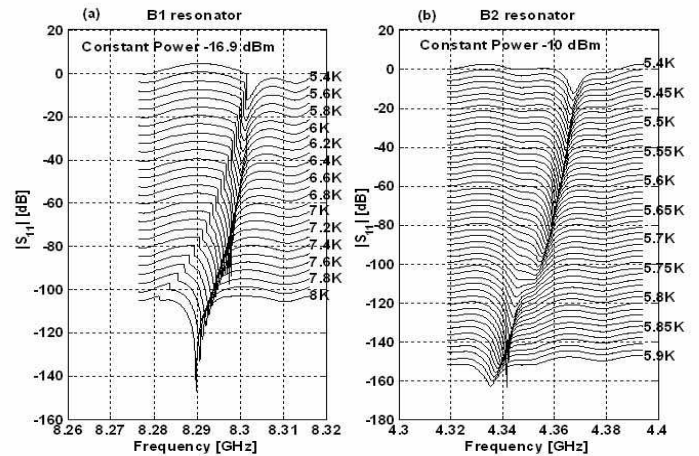}%
\caption{Subplot (a) exhibits the nonlinear resonance frequency response of B1
measured under constant input power of $-16.9$ dBm while increasing the
temperature from $5.4\mathrm{K}$ through $8\mathrm{K}$ in steps of
$0.1\mathrm{K}$. In addition to the gradual resonance frequency shift,
one can notice the resonance evolution as the temperature increases. Subplot
(b) shows B2 second mode measured at a constant input power of $-10$ dBm while
increasing the temperature from $5.4\mathrm{K}$ through
$5.9\mathrm{K}$ in steps of $0.01\mathrm{K}$. In addition to the
gradual resonance frequency shift, one can notice a temperature induced
critical coupling occurring at $5.8\mathrm{K}$. The resonance curves
were shifted vertically by a constant offset for clarity.}%
\label{bothrestemp}%
\end{center}
\end{figure}

The physical properties of superconductors such as London penetration depth,
surface resistance and kinetic inductance are strongly dependent on
temperature \cite{Sheen}. Whereas temperature varying measurements are
generally used in the characterization process of superconductors and in the
determination of London penetration depth \cite{London}, we vary the
temperature here in order to examine its effect on the nonlinear dynamics of
the resonators. The measurements were performed at temperature ranges well
below T$_{c}$ of the resonators, thus the exhibited nonlinear dynamics
presented here are features of the superconducting phase only. In Fig.
\ref{bothrestemp} subplot (a) we show a $S_{11}$ parameter amplitude
measurement versus frequency of B1 resonator attained at its third resonance
mode \cite{nonlinear features BB}. The input power was set to $-16.9$ dBm, a
power level at which the bifurcation is already present in the resonance
curve, afterwards the temperature was increased from $5.4%
%TCIMACRO{\unit{K}}%
%BeginExpansion
\mathrm{K}%
%EndExpansion
$ through $8%
%TCIMACRO{\unit{K}}%
%BeginExpansion
\mathrm{K}%
%EndExpansion
$ in steps of $0.1%
%TCIMACRO{\unit{K}}%
%BeginExpansion
\mathrm{K}%
%EndExpansion
.$ The nonlinear dynamics of the resonance curve as the temperature is
increased exhibit three main nonlinear features, the resonance frequency
shifts towards lower frequencies, the bifurcation step decreases and the
bifurcation frequency becomes lower. Whereas the first nonlinear feature can
be explained in terms of increase in the resonator inductance per unit length
$L$ due to temperature rise \cite{Sheen}, the second and third features are
less straightforward, and are more likely related to the effect of temperature
on the underlying nonlinear mechanism responsible for the observed effects.
Likewise in subplot (b), B2 resonator at its second mode exhibits a similar
behavior, though the input power applied in this case is $-10$ dBm, which is
lower than the bifurcation threshold of this resonance ( $\sim$ $-9$ dBm).
Similarly the resonance frequency shifts gradually towards lower frequencies
as the ambient temperature is varied from $5.4%
%TCIMACRO{\unit{K}}%
%BeginExpansion
\mathrm{K}%
%EndExpansion
$ through $5.9%
%TCIMACRO{\unit{K}}%
%BeginExpansion
\mathrm{K}%
%EndExpansion
$ in steps of $0.01%
%TCIMACRO{\unit{K}}%
%BeginExpansion
\mathrm{K}%
%EndExpansion
$. Moreover at $5.8%
%TCIMACRO{\unit{K}}%
%BeginExpansion
\mathrm{K}%
%EndExpansion
,$ the resonator even achieves critical coupling condition ($S_{11}\simeq0$)
via temperature varying only.%

%TCIMACRO{\FRAME{ftbpFU}{3.8623in}{2.5512in}{0pt}{\Qcb{Subplot (a) exhibits the
%nonlinear resonance frequency response of B1 measured under constant input
%power of $-16.9$ dBm while increasing the temperature from $5.4\unit{K}$
%through $8\unit{K}$ in steps of $0.1\unit{K}$. In addition to the gradual
%resonance frequency shift, one can notice the resonance evolution as the
%temperature increases. Subplot (b) shows B2 second mode measured at a constant
%input power of $-10$ dBm while increasing the temperature from $5.4\unit{K}$
%through $5.9\unit{K}$ in steps of $0.01\unit{K}$. In addition to the gradual
%resonance frequency shift, one can notice a temperature induced critical
%coupling occurring at $5.8\unit{K}$. The resonance curves were shifted
%vertically by a constant offset for clarity.}}{\Qlb{bothrestemp}%
%}{bothrestemp.eps}{\special{ language "Scientific Word";  type "GRAPHIC";
%display "USEDEF";  valid_file "F";  width 3.8623in;  height 2.5512in;
%depth 0pt;  original-width 7.4184in;  original-height 5.6334in;
%cropleft "0";  croptop "1";  cropright "1";  cropbottom "0";
%filename 'bothrestemp.eps';file-properties "XNPEU";}}}%
%BeginExpansion
%EndExpansion

\subsection{MAGNETIC FIELD DEPENDENCE}

\begin{figure}
[ptb]
\begin{center}
\includegraphics[
height=2.1958in,
width=3.3235in
]%
{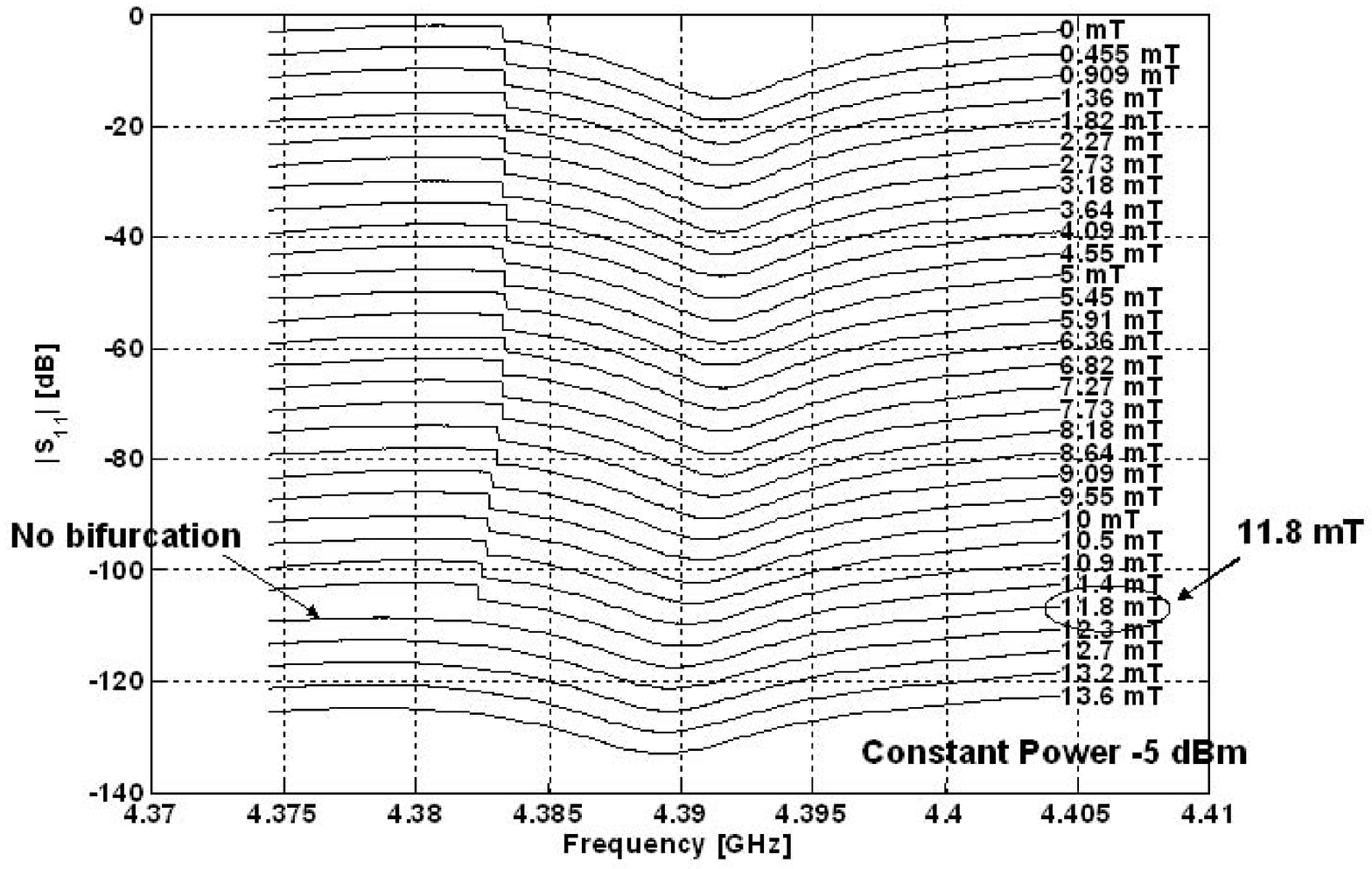}%
\caption{Increasing the magnetic field gradually from zero causes the
bifurcation in the B2 resonance band to disappear at relatively low value of
$11.8\mathrm{mT},$ while applying a constant input power level of $-5$
dBm. This bifurcation vanishing indicates that the bifurcation mechanism is
sensitive to magnetic field. The measured curves were shifted vertically by a
constant offset for clarity.}%
\label{magnojump}%
\end{center}
\end{figure}

\begin{figure}
[ptb]
\begin{center}
\includegraphics[
height=3.0735in,
width=3.4091in
]%
{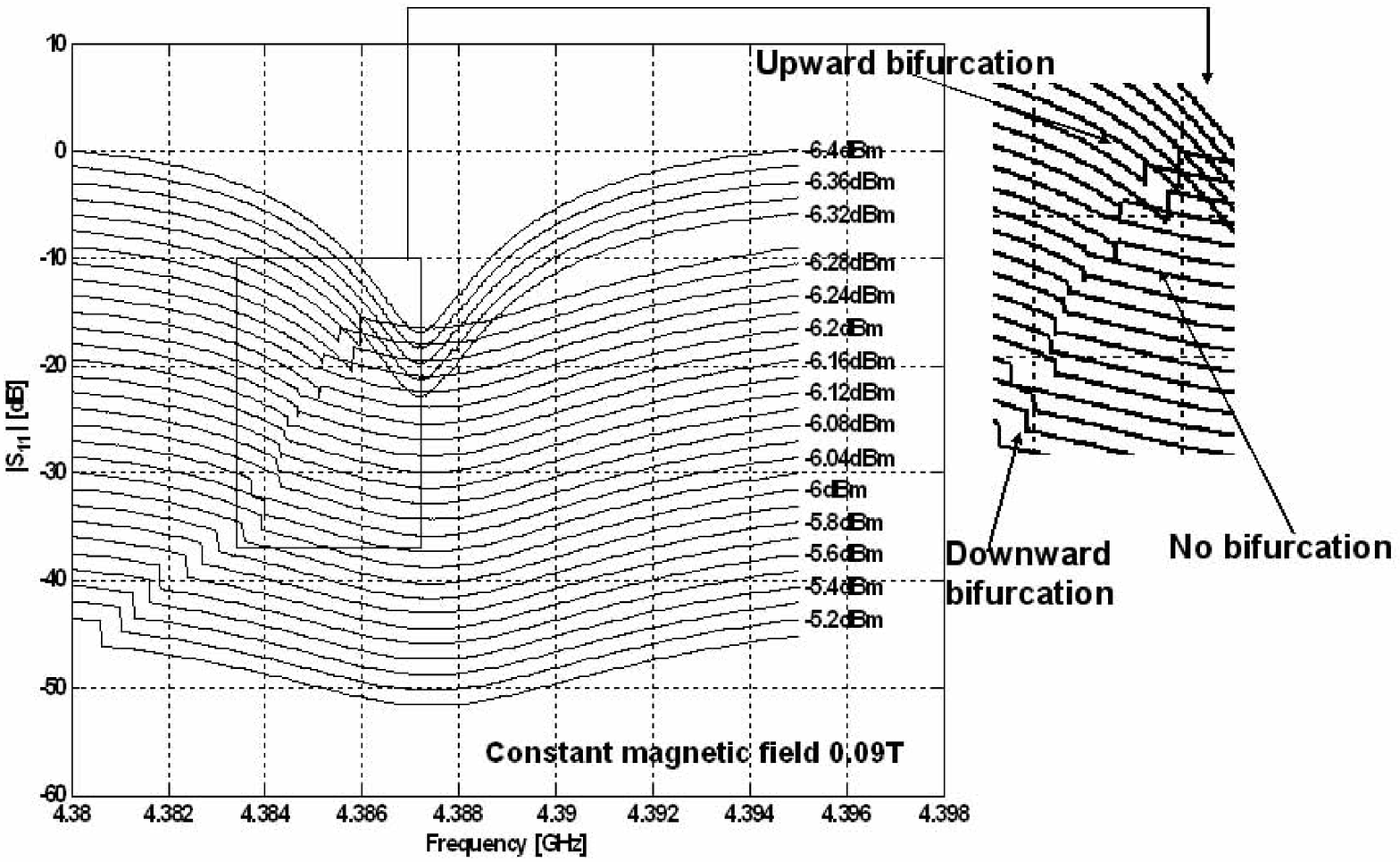}%
\caption{B2 nonlinear resonance response measured under a constant magnetic
field of $0.09\mathrm{T},$ while increasing the input power. The figure
exhibits additional dependencies of the resonance response curves on applied
magnetic field. The resonance which starts as a Lorentzian develops into a
resonance curve having an upward bifurcation as the power is increased,
afterwards into a curve with no bifurcation, followed by a resonance curve
having a downward bifurcation as the input power is increased further. The
measured curves were shifted vertically by a constant offset for clarity.}%
\label{b2s11magconst}%
\end{center}
\end{figure}

In order to better examine the extrinsic origin of the observed
nonlinearities, we have investigated the dependence of the nonlinear
resonances on magnetic field. The magnetic field dependence was measured by
applying two methods, one by setting a constant input power and varying the
magnetic field, and two by setting a constant magnetic field and varying the
input power. The magnetic field applied was perpendicular to the resonator
plane. The magnetic field measurements exhibiting B2 second resonance mode are
summarized in Figs. \ref{magnojump}, \ref{b2s11magconst}. In order to obtain
the results of Fig. \ref{magnojump} we have applied a constant input power of
$-5$ dBm to the resonator, and measured its $S_{11}$ response while increasing
the applied magnetic field by small steps. Above some low magnetic field
threshold of $11.8%
%TCIMACRO{\unit{mT}}%
%BeginExpansion
\mathrm{mT}%
%EndExpansion
$, the left side bifurcation vanishes, thus suggesting that the physical
mechanism causing the bifurcations is sensitive to low magnetic field. In Fig.
\ref{b2s11magconst} we applied the second method, where we set the magnetic
field to a constant level of $0.09%
%TCIMACRO{\unit{T}}%
%BeginExpansion
\mathrm{T}%
%EndExpansion
$ and measured the resonance response while increasing the input power. The
dynamic behavior measured using this method, as the input power is increased,
undergoes through the following sequential phases, symmetrical and Lorentzian
curves, resonance curves containing an upward bifurcation, a curve without
bifurcation, and finally resonance curves containing a downward bifurcation.%

%TCIMACRO{\FRAME{ftbpFU}{3.3235in}{2.1958in}{0pt}{\Qcb{Increasing the magnetic
%field gradually from zero causes the bifurcation in the B2 resonance band to
%disappear at relatively low value of $11.8\unit{mT},$ while applying a
%constant input power level of $-5$ dBm. This bifurcation vanishing indicates
%that the bifurcation mechanism is sensitive to magnetic field. The measured
%curves were shifted vertically by a constant offset for clarity.}}%
%{\Qlb{magnojump}}{magnojump.eps}{\special{ language "Scientific Word";
%type "GRAPHIC";  display "USEDEF";  valid_file "F";  width 3.3235in;
%height 2.1958in;  depth 0pt;  original-width 10.6666in;
%original-height 6.8822in;  cropleft "0";  croptop "1";  cropright "1";
%cropbottom "0";  filename 'magnojump.eps';file-properties "XNPEU";}}}%
%BeginExpansion
%EndExpansion
%

%TCIMACRO{\FRAME{ftbpFU}{3.4091in}{3.0735in}{0pt}{\Qcb{B2 nonlinear resonance
%response measured under a constant magnetic field of $0.09\unit{T},$ while
%increasing the input power. The figure exhibits additional dependencies of the
%resonance response curves on applied magnetic field. The resonance which
%starts as a Lorentzian develops into a resonance curve having an upward
%bifurcation as the power is increased, afterwards into a curve with no
%bifurcation, followed by a resonance curve having a downward bifurcation as
%the input power is increased further. The measured curves were shifted
%vertically by a constant offset for clarity.}}{\Qlb{b2s11magconst}%
%}{b2s11magconst.eps}{\special{ language "Scientific Word";  type "GRAPHIC";
%display "USEDEF";  valid_file "F";  width 3.4091in;  height 3.0735in;
%depth 0pt;  original-width 14.9474in;  original-height 9.5838in;
%cropleft "0";  croptop "1";  cropright "1";  cropbottom "0";
%filename 'b2s11magconst.eps';file-properties "XNPEU";}}}%
%BeginExpansion
%EndExpansion

\subsection{FREQUENCY SWEEP TIME ANALYSIS}

\begin{figure}
[ptb]
\begin{center}
\includegraphics[
height=1.689in,
width=3.4324in
]%
{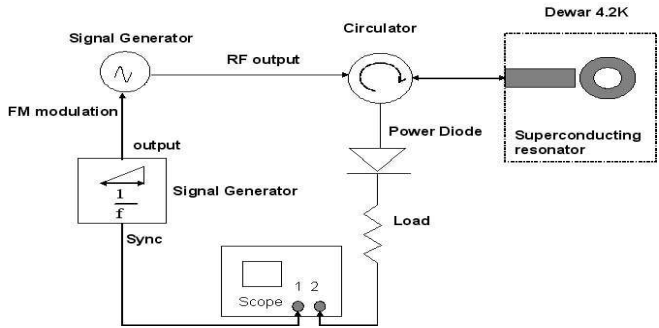}%
\caption{Frequency sweep time analysis setup. The frequency sweep time of the
microwave signal generator was FM modulated by a saw tooth waveform with
frequency $f$. The reflected power from the resonator was measured by a power
diode and oscilloscope.}%
\label{fmsetup}%
\end{center}
\end{figure}

\begin{figure}
[ptb]
\begin{center}
\includegraphics[
height=2.0825in,
width=3.4065in
]%
{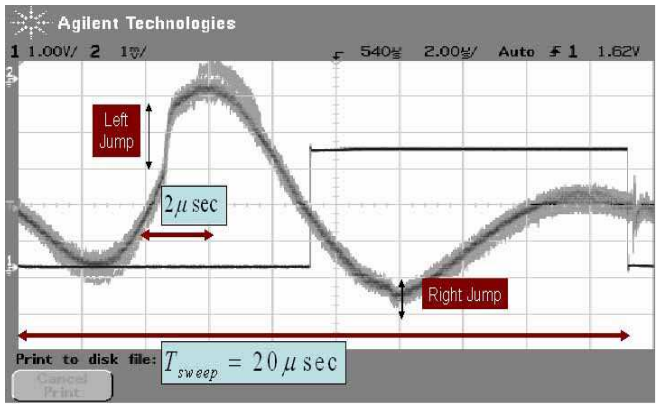}%
\caption{Frequency sweep time measurement. The figure displays the resonance
measured by Agilent oscilloscope while applying a saw tooth FM modulation of
frequency $50\mathrm{kHz}$ ( $T_{sweep}=$20$\mathrm{\mu s}$) to
the signal generator. The left and right bifurcations of the resonance are
still apparent in spite of the fast rate frequency sweep. Thus indicating that
the bifurcations do not originate from any global heating mechanism.}%
\label{timesweep}%
\end{center}
\end{figure}

Resistive losses and heating effects are typically characterized by relatively
long time scales \cite{Thermally induced nonlinear behaviour,power dependent
effects observed for sc stripline resonator}. In attempt to consider whether
such effects are responsible for the observed nonlinearities in general and
for the bifurcations in particular, we have run frequency sweep time analysis
using the experimental setup depicted in Fig. \ref{fmsetup}. We have
controlled the frequency sweep cycle of an Anritsu model signal generator via
FM modulation. The FM modulation was obtained by feeding the Anritsu with a
saw tooth waveform generated by another signal generator having $1/f$ sweep
time cycle. The reflected power from the resonator was redirected using a
circulator and measured by a power diode and oscilloscope. The left and right
hand bifurcations of B2 $\sim4.39%
%TCIMACRO{\unit{GHz}}%
%BeginExpansion
\mathrm{GHz}%
%EndExpansion
$ resonance were measured using this setup, while applying increasing FM
modulation frequencies: $1,10,20,30,40,50,100,150$ and $200%
%TCIMACRO{\unit{kHz}}%
%BeginExpansion
\mathrm{kHz}%
%EndExpansion
$. In Fig. \ref{timesweep} we present a measurement result obtained at $50%
%TCIMACRO{\unit{kHz}}%
%BeginExpansion
\mathrm{kHz}%
%EndExpansion
$ FM modulation, or alternatively $T_{sweep}$ of $20%
%TCIMACRO{\unit{\U{3bc}s}}%
%BeginExpansion
\mathrm{\mu s}%
%EndExpansion
$. The FM modulation applied was $\pm20%
%TCIMACRO{\unit{MHz}}%
%BeginExpansion
\mathrm{MHz}%
%EndExpansion
$ around $4.4022%
%TCIMACRO{\unit{GHz}}%
%BeginExpansion
\mathrm{GHz}%
%EndExpansion
$ center frequency. The measured resonance response appears inverted in the
figure due to the negative output polarity of the power diode. The fact that
both bifurcations continue to occur within the resonance lineshape (see Fig.
\ref{timesweep}), in spite of the short duty cycles that are on the order of
$\sim%
%TCIMACRO{\unit{\U{3bc}s}}%
%BeginExpansion
\mathrm{\mu s}%
%EndExpansion
$, indicates that heating processes which have typical time scale on the order
of $%
%TCIMACRO{\unit{s}}%
%BeginExpansion
\mathrm{s}%
%EndExpansion
$ to $%
%TCIMACRO{\unit{ms}}%
%BeginExpansion
\mathrm{ms}%
%EndExpansion
$ \cite{Thermally induced nonlinear behaviour} are unlikely to cause these
effects. Though there are also few who reported a smaller time constant of
about $3%
%TCIMACRO{\unit{\U{3bc}s}}%
%BeginExpansion
\mathrm{\mu s}%
%EndExpansion
$ \cite{Thermally-induced nonlinearities surface impedance sc YBCO} as a
result of local heating effects.

In order to obtain an estimated low limit of local heating time scales in our
NbN films, we apply a simple hot spot heating model to the resonators
\cite{Bol,Nonequilibrium,Use}. By further assuming that the substrate is
isothermal and that the hot spot is dissipated mainly down into the substrate
rather than along the film \cite{Bol}, one can evaluate the characteristic
relaxation time of the hot spot using the equation $\tau=Cd/\alpha,$ where $C$
is the heat capacity of the superconducting film (per unit volume), $d$ is the
film thickness, and $\alpha$ is the thermal surface conductance between the
film and the substrate \cite{Nonequilibrium}. Substituting for our B2 NbN
resonator yields a characteristic relaxation time of $\tau\simeq
5.4\cdot10^{-8}%
%TCIMACRO{\unit{s}}%
%BeginExpansion
\mathrm{s}%
%EndExpansion
,$ where the parameters $C\simeq2.7\cdot10^{-3}%
%TCIMACRO{\unit{W}}%
%BeginExpansion
\mathrm{W}%
%EndExpansion%
%TCIMACRO{\unit{cm}}%
%BeginExpansion
\mathrm{cm}%
%EndExpansion
^{-3}%
%TCIMACRO{\unit{K}}%
%BeginExpansion
\mathrm{K}%
%EndExpansion
^{-1}$ (NbN) \cite{Use}, $d=3000%
%TCIMACRO{\unit{\U{212b}}}%
%BeginExpansion
\mathrm{\text{\AA}}%
%EndExpansion
$ (B2 thickness)$,$ and $\alpha\simeq1.5%
%TCIMACRO{\unit{W}}%
%BeginExpansion
\mathrm{W}%
%EndExpansion%
%TCIMACRO{\unit{cm}}%
%BeginExpansion
\mathrm{cm}%
%EndExpansion
^{-2}%
%TCIMACRO{\unit{K}}%
%BeginExpansion
\mathrm{K}%
%EndExpansion
^{-1}$ at $4.2%
%TCIMACRO{\unit{K}}%
%BeginExpansion
\mathrm{K}%
%EndExpansion
$ (sapphire substrate) \cite{Use}, have been used. Similar calculation based
on values given in Ref. \cite{Bol} yields $\tau\simeq2.1\cdot0^{-9}%
%TCIMACRO{\unit{s}}%
%BeginExpansion
\mathrm{s}%
%EndExpansion
.$ These time scales are of course 2-3 orders of magnitude lower than the time
scales examined by the FM modulation setup.

Nevertheless intermodulation products observed in these resonators \cite{IMD
amplifier}, which generally require fast nonlinear response for their
generation$\ $on the order of $\leq10^{-10}%
%TCIMACRO{\unit{s}}%
%BeginExpansion
\mathrm{s}%
%EndExpansion
$ \cite{nonlinear dynamics}, further excludes heating effects, though not
entirely those associated with very fast heating mechanisms similar to hot
spots for example.%

%TCIMACRO{\FRAME{ftbpFU}{3.4324in}{1.689in}{0pt}{\Qcb{Frequency sweep time
%analysis setup. The frequency sweep time of the microwave signal generator was
%FM modulated by a saw tooth waveform with frequency $f$. The reflected power
%from the resonator was measured by a power diode and oscilloscope.}%
%}{\Qlb{fmsetup}}{fmsetup.eps}{\special{ language "Scientific Word";
%type "GRAPHIC";  display "USEDEF";  valid_file "F";  width 3.4324in;
%height 1.689in;  depth 0pt;  original-width 12.2112in;
%original-height 7.5161in;  cropleft "0";  croptop "1";  cropright "1";
%cropbottom "0";  filename 'fmsetup.eps';file-properties "XNPEU";}}}%
%BeginExpansion
%EndExpansion
%

%TCIMACRO{\FRAME{ftbpFU}{3.4065in}{2.0825in}{0pt}{\Qcb{Frequency sweep time
%measurement. The figure displays the resonance measured by Agilent
%oscilloscope while applying a saw tooth FM modulation of frequency
%$50\unit{kHz}$ ( $T_{sweep}=$20$\unit{\U{3bc}s}$) to the signal generator. The
%left and right bifurcations of the resonance are still apparent in spite of
%the fast rate frequency sweep. Thus indicating that the bifurcations do not
%originate from any global heating mechanism.}}{\Qlb{timesweep}}{timesweep.eps}%
%{\special{ language "Scientific Word";  type "GRAPHIC";  display "USEDEF";
%valid_file "F";  width 3.4065in;  height 2.0825in;  depth 0pt;
%original-width 10.0145in;  original-height 4.8326in;  cropleft "0";
%filename 'timesweep.eps';file-properties "XNPEU";}}}%
%BeginExpansion
%EndExpansion

\subsection{COLUMNAR STRUCTURE}

\begin{figure}
[ptb]
\begin{center}
\includegraphics[
height=2.2148in,
width=3.3944in
]%
{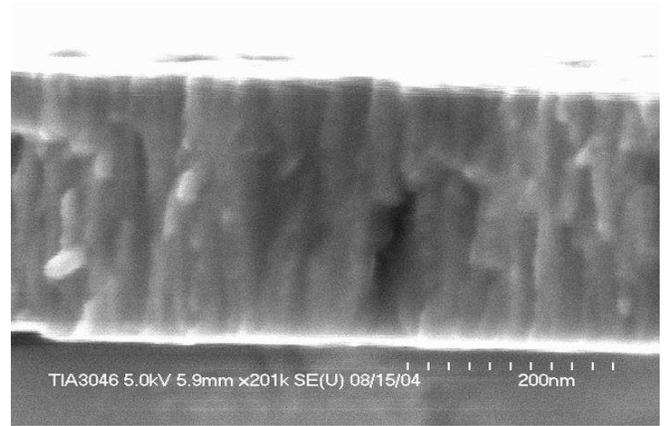}%
\caption{SEM micrograph displaying a $2200\mathrm{\text{\AA}}$ NbN film
deposited on a thin sapphire substrate using similar sputtering conditions as
B2 resonator. The micrograph exhibits clearly the columnar structure of the
NbN film and its grain boundaries.}%
\label{sem}%
\end{center}
\end{figure}

It is well known from numerous research works done in the past \cite{rf
superconducting properties of reactively sputtered NbN,superconducting
properties and NbN structure}, that NbN films can grow in a granular columnar
structure under certain deposition conditions. Such columnar structure may
even promote the growth of random built-in Josephson junctions at the grain
boundaries of the NbN films. To study the NbN granular structure we have
sputtered about $2200%
%TCIMACRO{\unit{\U{212b}}}%
%BeginExpansion
\mathrm{\text{\AA}}%
%EndExpansion
$ thickness NbN film on a thin small rectangular sapphire substrate of $0.2$ $%
%TCIMACRO{\unit{mm}}%
%BeginExpansion
\mathrm{mm}%
%EndExpansion
$ thickness. The sputtering conditions applied were similar to those used in
the fabrication of B2 resonator. Following the sputtering process, the thin
sapphire was cleaved, and a scanning electron microscope (SEM) micrograph was
taken at the cleavage plane. The SEM micrograph in Fig. \ref{sem} which
clearly shows the columnar structure of the deposited NbN film and its grain
boundaries, further supports our weak link hypothesis. The typical diameter of
each NbN column is about $20$ $%
%TCIMACRO{\unit{nm}}%
%BeginExpansion
\mathrm{nm}%
%EndExpansion
.$%

%TCIMACRO{\FRAME{ftbpFU}{3.3944in}{2.2148in}{0pt}{\Qcb{SEM micrograph
%displaying a $2200\unit{\U{212b}}$ NbN film deposited on a thin sapphire
%substrate using similar sputtering conditions as B2 resonator. The micrograph
%exhibits clearly the columnar structure of the NbN film and its grain
%boundaries.}}{\Qlb{sem}}{sem.eps}{\special{ language "Scientific Word";
%type "GRAPHIC";  display "USEDEF";  valid_file "F";  width 3.3944in;
%height 2.2148in;  depth 0pt;  original-width 10.0024in;
%original-height 7.4936in;  cropleft "0";  croptop "1";  cropright "1";
%cropbottom "0";  filename 'sem.eps';file-properties "XNPEU";}}}%
%BeginExpansion
%EndExpansion

\section{DISCUSSION}

As we argue herein the physical mechanism that may potentially account for the
nonlinear dynamics of these NbN resonators, is the extrinsic weak link
mechanism, or more specifically Josephson junctions forming at the columnar
structure boundaries of the NbN films. In addition, we show that this
hypothesis is qualitatively consistent with some of our measurement results.

The measurement results which support the Josephson junction hypothesis can be
summarized as follows:

(a) The columnar structure of the NbN films and their grain boundaries
demonstrated in Fig. \ref{sem}, which may allow the formation of Josephson
junctions between the NbN grains.

(b) The onset of bifurcations and nonlinear features occurring at considerably
low input power levels, about 2-4 orders of magnitude lower than the onset of
Duffing nonlinearity in the Nb resonator ( $\sim$ $-28$ in Fig. \ref{b1s11f1}
versus $\sim$ $10$ dBm in Fig. \ref{nb_duffing}). This low power threshold
highly implies an extrinsic nonlinear origin. Such extrinsic nonlinearity
could be ac driven Josephson junctions, having low onset of instability
amplitude, or junctions characterized by a low Josephson critical current,
resulting in bifurcations in the dynamics of the gauge invariant phase
difference across the Josephson junctions, at relatively low ac current
amplitudes \cite{Clem}.

(c) The bifurcation dependence on magnetic field as was shown in Fig.
\ref{magnojump}, where B2 resonance bifurcation vanished as the perpendicular
magnetic field was increased above some low magnetic field threshold
$\sim11.8$ $%
%TCIMACRO{\unit{mT}}%
%BeginExpansion
\mathrm{mT}%
%EndExpansion
$, which is about $3.5$ times lower than $H_{c_{1}}$ of NbN reported for
example in \cite{nonlinear dynamics}. On the other hand Josephson junctions
are known for their sensitive dependence on magnetic field.

(d) The observation of a very small number of bifurcations in the resonance
lineshape, (two in Figs. \ref{b1s11f1}, \ref{hysteresis} and three in Fig.
\ref{b3fb}), compared to the large number of built in Josephson junctions
expected according to the columnar structure model, may be explained
qualitatively by one of the following possible scenarios. The coupling between
the Josephson junctions and the global resonator activates only a small number
of Josephson junctions characterized by some limiting properties such as
Josephson resonance frequency falling within the resonance band of the global
resonator, or Josephson onset of instability amplitude being lower than the
driving oscillation amplitude inside the resonator. Whereas the other scenario
assumes a synchronization mechanism acting on the junctions, thus causing them
to lock and evolve with the same frequency. Such synchronization mechanism may
be similar to the mechanism recently analyzed in Ref. \cite{synchronization},
where synchronization between nonlinear oscillators is achieved through
nonlinear frequency pulling and reactive coupling.

(e) The rapid frequency sweep experiments, which excluded most heating effects
(except possible very fast local heating mechanisms), further supports the
Josephson junction hypothesis, since dynamic states of ac biased Josephson
junctions may involve very little dissipation (vanishing dc voltage).

Moreover it was shown in several experimental works published recently
\cite{opposed hammerhead,Pinch resonances in rf,nonlinear multilevel,Giant},
examining the resonance lineshape of a radio frequency resonators coupled to a
superconducting ring containing a single Josephson junction, that interesting
nonlinear dynamics can develop in the RF circuit resonance response as the
external magnetic flux applied through the ring is varied. Among the nonlinear
dynamics reported therein one can name, bifurcations in the resonance
lineshape, single and opposed fold bifurcations (hysteresis loops)
\cite{opposed hammerhead}, pinch resonances, where the opposed fold resonances
appear to touch (pinch off) \cite{Pinch resonances in rf}, and even stochastic
jumps \cite{nonlinear multilevel}. These experimental results are shown to be
solutions of the nonlinear equations of motion for the system. In addition,
different solutions of the nonlinear equation of Josephson junction driven by
ac field, for various limiting cases, can be found in Ref. \cite{Clem}.
Whereas the nonlinear dynamics of Josephson junctions incorporated in resonant
cavities, which was analyzed recently in several theoretical and experimental
studies, mainly for the purpose of developing sources of coherent microwave
radiation, can be found in Refs. \cite{Elmaas1,Elmaas2,Tornes,point
ms,Stimulated}, and in the cited references therein. Although these works do
not analyze the effect of Josephson junctions on the resonance response of the
cavity (a single mode is generally assumed), they provide a good understanding
of the I-V characteristics of these arrays.

In spite of the qualitative arguments presented here, a large part of the
underlying physics of these nonlinear resonators is still not fully
understood, and further theoretical work may be needed in order to account for
most of the experimental data reported here.

\begin{center}
\textbf{ACKNOWLEDGEMENTS}
\end{center}

B.A. would like to thank John R. Clem for bringing to his attention some
recent work done on Josephson junctions. E.B. would especially like to thank
Michael L. Roukes for supporting the early stage of this research and for many
helpful conversations and invaluable suggestions. Very helpful conversations
with Oded Gottlieb, Gad Koren, Emil Polturak, and Bernard Yurke are also
gratefully acknowledged. This work was supported by the German Israel
Foundation under grant 1-2038.1114.07, the Israel Science Foundation under
grant 1380021, the Deborah Foundation and Poznanski Foundation.

%Just because of unusual number of tables stacked at end
\bibliographystyle{plain}
\bibliography{apssamp}
%Produces the bibliography via BibTeX.%
%TCIMACRO{\TeXButton{TeX field}{\enlargethispage{-5in}}}%
%BeginExpansion
\enlargethispage{-5in}%
%EndExpansion

\end{document}